# Chirality Amplification and Detection by Tactoids of Lyotropic Chromonic Liquid Crystals†

Chenhui Peng[a] and Oleg D. Lavrentovich*[a]

**Abstract:** Detection of chiral molecules requires amplification of chirality to measurable levels. Typically, amplification mechanisms are considered at the microscopic scales of individual molecules and their aggregates. Here we demonstrate chirality amplification and visualization of structural handedness in water solutions of organic molecules that extends over the scale of many micrometers. The mechanism is rooted in the long-range elastic nature of orientational order in lyotropic chromonic liquid crystals (LCLCs) formed in water solutions of achiral disc-like molecules. The nematic LCLC coexists with its isotropic counterpart, forming elongated tactoids; spatial confinement causes structural twist even when the material is nonchiral. Minute quantities of chiral molecules such as amino acid L-alanine and limonene transform the racemic array of left- and right-twisted tactoids into a homochiral set. The left and right chiral enantiomers are readily distinguished from each other as the induced structural handedness is visualized through a simple polarizing microscope observation. The effect is important for developing our understanding of chirality amplification mechanisms; it also might open new possibilities in biosensing.

## Introduction

Living systems utilize chiral molecules with only one helical sense, such as L-amino acids and D-nucleotides. The asymmetry might have occurred as a chance creation of minute enantiomeric excess in a racemic mixture, but to raise it to the detectable macroscopic effects, one needs effective amplification mechanisms. The amplification of chirality has been considered in asymmetric autocatalysis[1, 2], covalent polymerization[3, 4], and non-covalent supramolecular assembly[5-7], crystallization[8], etc. Although it is difficult to formulate the general rules of chirality amplification[5], there is an emerging understanding that the efficient amplification involves, first of all, molecular and supra-molecular level features[9]. Among these is the ability of individual organic molecules to adopt an intrinsically chiral shape (for example, a propeller shape of disc-like molecules[5]) and the ability of direct molecular interactions such as covalent, hydrogen or $\pi-\pi$ interactions to transfer chirality from one molecule to another. Here we demonstrate a chirality amplification mechanism that is based on the long-range forces, extending over the scales of tens of micrometers, much larger than the nanometer scale of a single molecule and covalent molecular interactions. The mechanism is rooted in the long-range elastic nature of orientational order in lyotropic chromonic liquid crystals (LCLCs).

LCLC molecules are plank-like with aromatic cores and hydrophilic ionic groups at the peripheries. Once dissolved in water, they self-assemble into aggregates, attracting each other face-to-face by weak noncovalent $\pi-\pi$ interactions[10-14]. When concentration increases, the aggregates elongate and eventually form the nematic phase by aligning parallel to each other. The nematic phase coexists with the isotropic phase in a broad range of concentrations and temperatures, typically in the form of spindle-like droplets called tactoids, observed in many lyotropic systems[15, 16], including LCLCs[17, 18]. The director configuration and the overall shape of tactoids are determined by the balance of the intrinsic elastic and anisotropic surface energies[17]. Because of the unusually small twist elastic constants in LCLCs[19, 20], the director field inside the tactoids assumes a twisted structure[17]. Such a structure is optically active. Since the LCLC is not chiral, the number of tactoids with the left and right twist in any given sample is approximately the same[17], i.e., the system is racemic, Fig.1.

The main result of this work is that a minute enantiomeric excess transforms the racemic array of tactoids into a macroscopically chiral system with a strong amplification of chirality. We demonstrate that a small amount of chiral molecules (R- and S-limonene, amino acid L-alanine, brucine) turns all or almost all of the tactoids into either right- or left-twisted structures, depending on the chirality of the additives, Fig.1. Chirality state of the enantiomers is readily visualized because the tactoids of opposite handedness exhibit different polarizing microscopy textures. The role of the chiral molecules is to tip the balance between the two types of twisted tactoids towards one type of structural handedness. Chirality amplification in the system of tactoids is characterized by an increase of optical activity by a factor of $10^3 - 10^4$, when compared to the optical activity of the isotropic solutions of the same composition. The effect illustrates yet another potential mechanism of homochirality of biological systems[21] and offer piotential applications in biosensing, chirality discrimination and enantioselective detection.

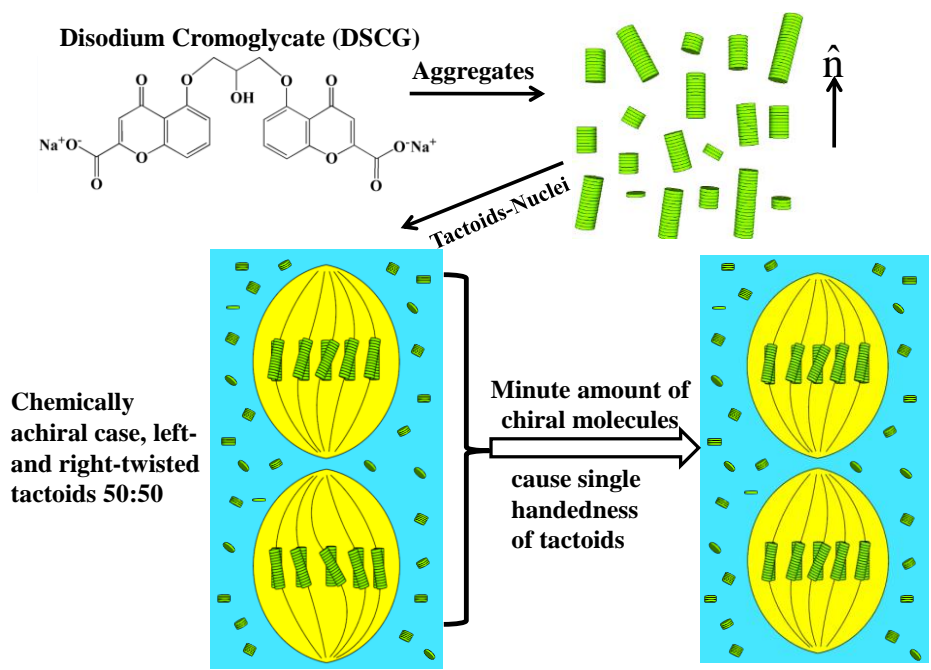

**Fig. 1** Scheme of chirality amplification in LCLC tactoids. The achiral LCLC disodium cromoglycate (DSCG) molecules self-assemble into elongated aggregates in water. The aggregates form nematic tactoids surrounded by the isotropic phase. The director at the footprint of the tactoid is mostly parallel to the axis that connects the two poles. As one moves towards the apex of the tactoid, the director twists to the right or to the left. If there are no chiral molecules, the left- and right-twisted tactoids are met in proportion of 50:50. Addition of chiral molecules eliminates parity, forcing the tactoids to be of the same handedness.

## Experimental

### Materials and samples

Disodium cromoglycate (DSCG) of purity 98% was purchased from Spectrum and used without further purification. We prepared 0.34 mol/kg (15 wt%) DSCG solutions in deionised water. Condensing agent poly(ethyleneglycole) (PEG) was introduced to the solution (concentration 0.8wt%, molecular weight 3350 g/mol) to stabilize the tactoids[17]. The biphasic region of coexisting nematic and isotropic phase in the DSCG+PEG solution extends from 19°C to 40°C; all experiments were performed at room temperature. We also explored DSCG+PEG solutions of the same composition, to which a chiral dopant was added: R-limonene (concentration $c = 1$ wt% with respect to the total weight of the sample), S-limonene ($c =1$ wt%), L-alanine ($c =0.1$ wt%), or brucine ($c =0.2$ wt%), Fig.2. All chiral additives were purchased from Sigma-Aldrich. Lower concentrations of L-alanine were also studied. All the mixtures were filled into flat glass cells by capillary forces and sealed by an epoxy glue. Glass substrates were coated with the polyimide SE-7511L (Nissan Chemical Industries, Ltd.) and rubbed unidirectionally to align the axes of tactoids, as their bases are located at the glass plates. The cell thickness was set by Mylar films (3M) at $180\,\mu m$.

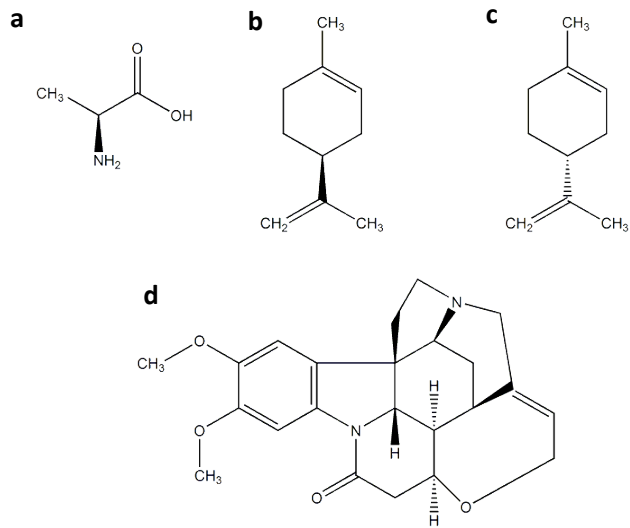

**Fig. 2** Molecular formulae of chiral additives. **a,** L-alanine. **b,** S-limonene. **c,** R-limonene. **d,** Brucine.

**Twist angle measurements**

The local orientation of the chromonic aggregates is described by the director $\hat{n}$ which is also the optic axis. A twisted director results in rotation of transmitted light polarization; this effect can be used to determine the twist angle $\tau$, i.e. the difference in azimuthal orientation of the director at the centre of the tactoid's footprint at the glass plate, $z=0$, and at its top, $z=d$, where $d$ is the tactoid's height (distance to the apex), Fig.1. To characterize the twist, we used the polarizing optical microscope Nikon Optiphot2-POL equipped with a Nikon P100S microspectrophotometer[22]. The angle $\gamma$ between the polarizer and analyser is continuously adjustable from 0 to $180°$. The polarizer is directed along the short axis of the tactoid. When the twist is relatively weak, the optical transmittance $T$ through the central part of the tactoid is described in the so-called Mauguin approximation as[23]

$$T = \cos^2\beta - (\tau/(2\delta))\sin(2\delta)\cos(2\beta)[(\tau/\delta)\tan\delta + \tan 2\beta] \quad (1)$$

where $\delta = \sqrt{\tau^2 + (\pi\Delta n d/\lambda)^2}$, $\beta = \gamma - \tau$, $\lambda = 550$ nm is the wavelength of light, $\Delta n = -0.026$ is the birefringence of the DSCG+PEG solution[17], $d$ is measured by fluorescence confocal polarizing microscope (FCPM)[17]. If the director does not twist, $\tau = 0$, the central part of tactoid is dark when the polarizers are crossed, $\gamma = 90°$. For twisted tactoids, the central part of the tactoid becomes dark when $\gamma$ is different from $90°$. By measuring the transmitted light intensity and fitting it with Eq. (1), one obtains the twist angle $\tau$ that is a measure of optical activity of the tactoid.

Presence of chiral molecules causes optical activity even in the isotropic phases[24]. It is thus important to compare the optical activity of the nematic tactoids to the optical activity of the isotropic melt of the same composition. The latter was determined by two techniques. In the first, the optical activity was determined through measuring the laser beam transmittance (He-Ne laser, wavelength 633 nm) through the cell as a function of the angle between two polarizers[25]. In the second approach, we used circular dichroism (CD) absorption spectroscopy (see ESI†).

## Results and Discussions

In absence of chiral additives, tactoids of the LCLC nematic phase coexisting with the isotropic phase show a twisted director [17]. The twisted structure is caused by the fact that the strong splay and bend deformations of the director, induced by the curved nematic-isotropic interface and by the surface anchoring at this interface, are relaxed through twist deformations; the elastic constant $K_2$ of twist is much smaller than the elastic constants $K_1$ of splay and $K_3$ of bend[19, 20]. The twist of director causes rotation of polarization of the transmitted light by an angle $\tau$ that is assigned either a positive sign (right-twisted tactoid, or RT) or a negative one (left-twisted tactoid, or LT). The two types are readily distinguishable in the textures viewed under the polarizing microscope when the analyser is rotated with respect to its usual crossed position with the polarizer, making an angle either smaller or larger than $\gamma = 90°$ [26, 27] (Fig. 3a,b). Tactoids that show extinction of the central part at $\gamma = 80° \pm 1°$ are of the LT type, and tactoids with extinction at $\gamma = 100° \pm 1°$ are of the RT type, Fig.3c. The twist angle $\tau$ value, extracted from the extinction angle $\gamma$, depends on the director gradients and thus on the width $w$ of the tactoid. For example, $\tau = -11° \pm 1°$ for the LT of width of $w = 50\,\mu m$ and $\tau = 9° \pm 1°$ for the RT with $w = 45\,\mu m$ (Fig. 3c).

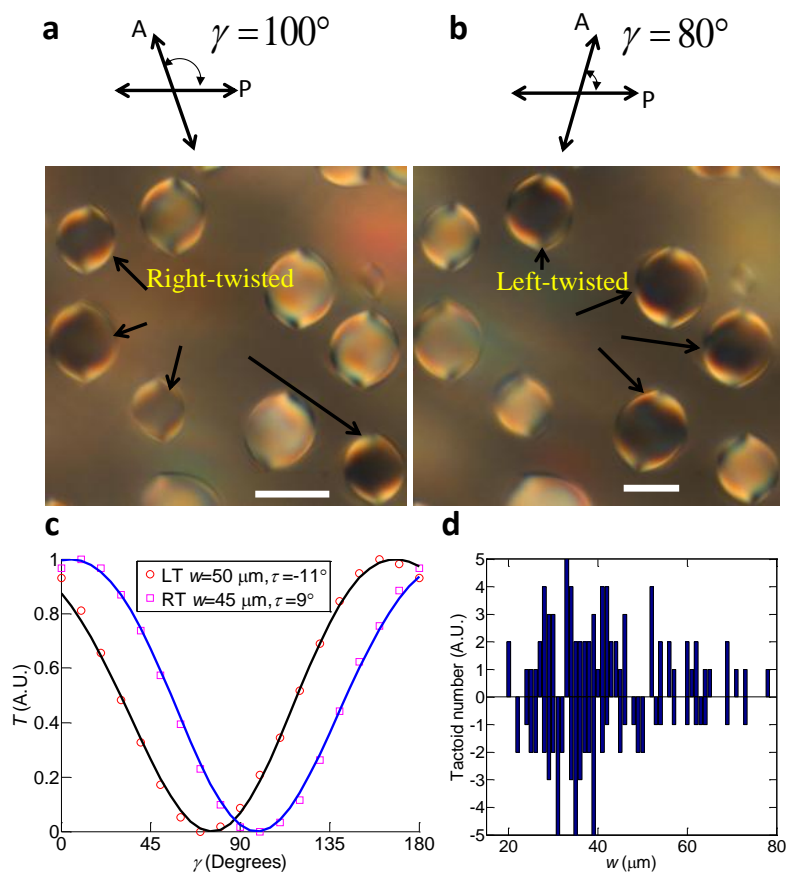

**Fig. 3** Polarising microscopy textures of tactoids and tactoids' handedness in DSCG+PEG water solutions with no chiral additives. **a-b**, Textures of left and right twisted tactoids are different when viewed with decrossed polarizer P and analyser A, $\gamma = 100°$ (**a**) and $\gamma = 80°$ (**b**). **c**, Transmitted intensity of polarized light as a function of the angle between the polarizer and analyser; the tactoid of width $50\,\mu m$ that appears dark at $\gamma = 80°$ have a twist angle of $\tau = -11° \pm 1°$ ; the tactoid of width $45\,\mu m$ $\tau = 9° \pm 1°$. **d**, Typical statistics of left- (negative numbers) and right- (positive numbers) twisted tactoids of different width in the racemic DSCG+PEG sample. Scale bar $50\,\mu m$.

Without chiral additives, the numbers $n_{LT}$ of LTs and $n_{RT}$ of RTs in a sufficiently large sample are equal to each other, Fig. 3d, i.e., the system is racemic. When the chiral material is added, the parity is broken and one type of the tactoids prevails, compare Fig.4a with RTs caused by adding R-limonene, with Fig.4b showing LTs in a LCLC doped with S-limonene. The effect is universal, as other chiral additives, such as L-alanine, Fig.4c, and brucine, Fig.4d, cause a similar effect of homochirality. The optical activity of tactoids in doped chiral system is somewhat higher than the optical activity of tactoids in the nematic (additive free) phase. For example, RTs and LTs of width $w = 45\,\mu m$ in presence of 1 wt% R- and S-limonene, show $\tau = 15° \pm 1°$ and $\tau = -15° \pm 1°$, respectively (Fig. 5a). Similar twist, $\tau = 15° \pm 1°$, is observed for RTs in DSCG+PEG doped with 0.2 wt% of brucine (Fig. 5c). L-alanine at 0.1 wt% produces $\tau = -12° \pm 1°$ in LTs of the same width $w = 45\,\mu m$ (Fig. 5b). The excess $(n_{LT} - n_{RT})/(n_{LT} + n_{RT})$ of one form of tactoids over the other as a function of L-alanine concentration is shown in Fig. 5d. The population of tactoids becomes practically homochiral already at concentrations as low as 0.04 wt%, Fig.5d.

The optical activity of tactoids with limonenes, defined as $\eta = |\tau|/cd$, is estimated as $\eta \approx 7.5° \times 10^5/(m \times wt\%)$ with the typical twist angle $15°$, tactoid's height $d = 20\,\mu m$ and chiral additive's concentration $c = 1\,wt\%$. It is of interest to compare $\eta$ to the optical activity $\eta_{iso}$ of isotropic solutions with the same chemical composition, obtained by a simple heating of the LCLC with chiral dopants into the isotropic phase. For an isotropic phase of DSCG+PEG with 1 wt% of R-limonene in a slab of thickness $180\,\mu m$, rotation angle of linearly polarized light at 633 nm is only about $0.02° \pm 0.003°$, so that $\eta_{iso}^{633} \approx 1.1° \times 10^2/(m \times wt\%)$. CD measurements in the ultraviolet part of the spectra (see ESI†) produce similar results. For the isotropic melts with chiral additives, the optical rotation angle is of the order of $0.01°$ for slabs of thickness $10\,\mu m$ in the spectral range (220-280) nm; for a longer wavelength 400 nm, the optical rotation angle drops to $0.0045°$ (see ESI†), so that $\eta_{iso}^{400} \approx 4.5° \times 10^2/(m \times wt\%)$. Hence, the optical activity of nematic tactoids is about 6800 times higher than that of the isotropic melt measured at 633 nm, , $\eta/\eta_{iso}^{633} \approx 6.8 \times 10^3$, and about 1600 times higher when $\eta_{iso}^{400}$ is measured at 400 nm, $\eta/\eta_{iso}^{400} \approx 1.6 \times 10^3$. Other materials show a similar or even stronger difference: L-alanine sample with $c = 0.1 wt\%$, shows $\eta \approx 6° \times 10^6/(m \times wt\%)$, while $\eta_{iso}^{633} \approx 8° \times 10^2/(m \times wt\%)$ and $\eta_{iso}^{400} \approx 2° \times 10^3/(m \times wt\%)$; therefore, $\eta/\eta_{iso}^{633} \approx 7.5 \times 10^3$ and $\eta/\eta_{iso}^{400} \approx 3 \times 10^3$. The reason for this large amplification power of the tactoids is that the chiral additive is not the sole agent responsible for the optical activity; the optical activity is already present because of the twist deformations inside the tactoids and the role of the chiral dopant is simply to shift the balance towards one of the two possible twisted forms.

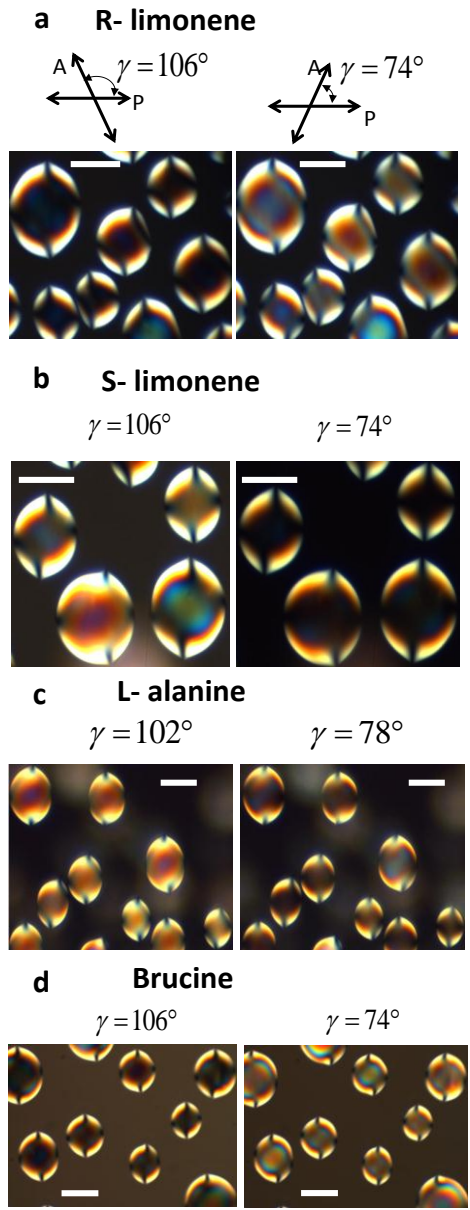

**Fig. 4** Textures of homochiral tactoids in a DSCG+PEG solutions with different chiral additives, viewed under a polarizing microscope with decrossed polarizers. **a**, R-limonene causes homochirality of the tactoidal array; all the tactoids exhibit dark centers when $\gamma = 106°$, **b**, S-limonene causes dark images of tactoids when $\gamma = 74°$. **c**, L-alanine causes extinction at $\gamma = 78°$. **d** Brucine causes extinction at $\gamma = 106°$. Scale bar $50 \mu m$.

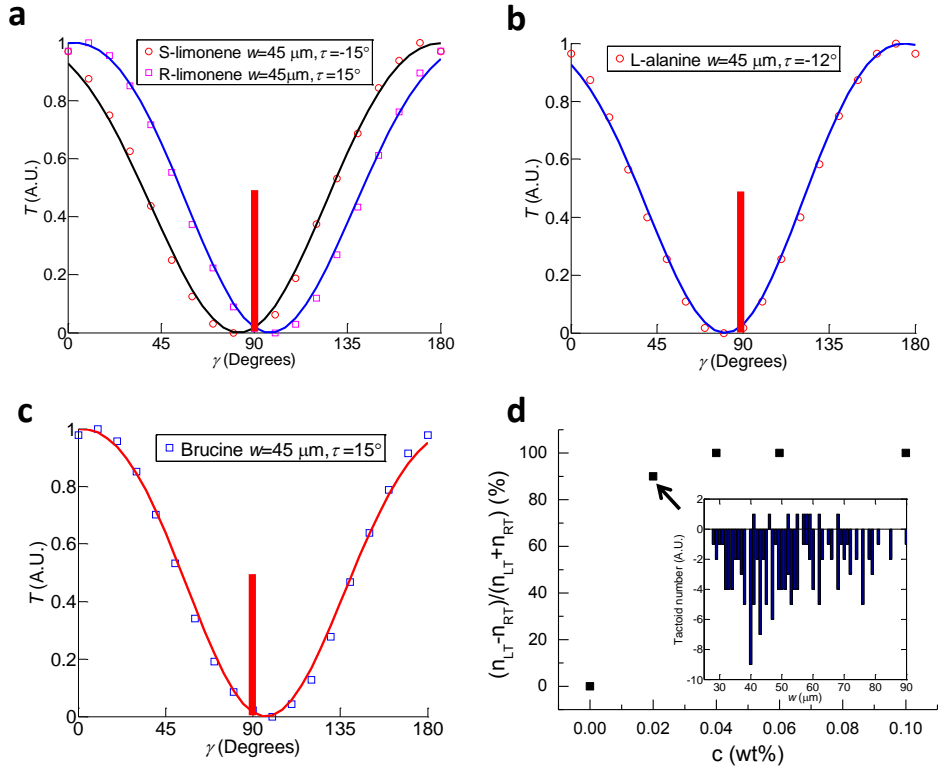

**Fig. 5** Optical transmittance of tactoids of width $45\,\mu m$ in DSCG+PEG solutions with chiral aditives. **a**, Tactoids exhibit a positive twist $\tau =15°$ when the sample contains 1 wt% of R-limonene and a negative twist $\tau =-15°$ in presence of 1wt% of S-limonene. **b**, L-alanine with $c=0.1\,\text{wt\%}$ causes $\tau =-12°$. **c**, Brucine, $c=0.2\,\text{wt\%}$, causes $\tau =15°$. **d**, Homochirality excess $(n_{LT}-n_{RT})/(n_{LT}+n_{RT})$ vs concentration $c$ of L-alanine in the DSCG+PEG solutions; inset shows the statistics of left-handed (negative numbers) vs right-handed (positive numbers) tactoids in the DSCG+PEG doped with 0.02wt% of L-alanine.

The optical activity of the tactoids can also be compared to that of LCLC in the homogeneous cholesteric phase (that exists at temperatures below the biphasic region). For this purpose, we explored 0.34 mol/kg (15 wt%) DSCG solutions in deionised water with added L-alanine but no PEG. In the homogeneous cholesteric phase that forms at the temperatures below $29°C$, the twist is caused solely by the chiral additive; there is no isotropic-tactoid interface that might cause an additional twist. The flat boundaries might only "unwind" the periodically twisted cholesteric, when the sample is too thin. The pitch $P$ of the homogeneous cholesteric was measured at different concentrations $c$ of L-alanine using fingerprint textures, Fig. 6. The slope of the dependency $1/P$ vs $c$ yields the helical twisting power (HTP), defined as $\bar{\eta}=1/cP$. For L-alanine in 15 wt% DSCG solution, we find $\bar{\eta}=8\times 10^3/(m\times \text{wt\%})$, comparable to the value obtained by McGinn et al[28]. Since $P$ is the distance over which the director rotates by $360°$, we deduce the effective optical activity of the homogeneous cholesteric, $\eta_{ch}\approx 2.9°\times 10^6/(m\times \text{wt\%})$. In contrast, the twisted tactoids doped with L-alanine show the optical activity $\eta\approx 6°\times 10^6/(m\times \text{wt\%})$ that is about two times stronger than that of the cholesteric. The effect is expected because the tactoids show twist even when there are no chiral dopants.

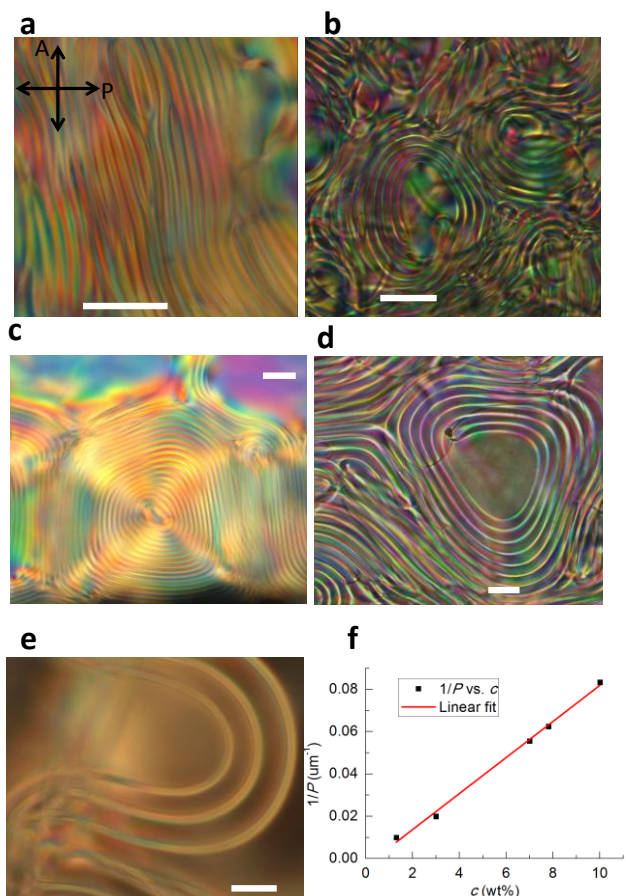

**Fig. 6** Pitch measurements for homogeneous cholesteric LCLC phase. **a-e**, Fingerprint textures in DSCG solution with L-alanine at the concentrations 10wt%, 7.8wt%, 7wt%, 3wt% and 1.3wt% respectively. The cell thickness ranges from 180 μm to 540 μm. **f**, Dependency $1/P$ vs. $c$. Scale bar 50 μm. P and A represent polarizer and analyzer.

## Discussion and Conclusion

In organic molecular systems, the mechanisms of chirality amplification[29] include "sergeants-and-soldiers" [30-33] and "majority-rule" principles[34-36]. The first refers to a situation when a small population of chiral units (sergeants) can induce the chirality of a large number of achiral units (soldiers). The second is applicable when a small enantiomeric excess in the racemic mixture of chiral units establishes homochirality. Both effects are dominated by the short range interactions of near neighbours[37]. In the LCLC tactoids, chirality amplification is based on the collective behaviour over the macroscopic length scales of confinement (microns and tens of microns). The added chiral dopants break parity in probability of finding left- and right-twisted tactoids and lead to homochirality of the tactoidal arrays.

The LCLC tactoids-based approach to amplification and detection of chiral additives has an important useful feature: it produces optical response that is directly associated with the enantiometric form of the chiral dopant. For example, tactoidal arrays with R-limonene and with S-limonene are readily distinguishable under the polarizing microscope with uncrossed polarisers since their textures are very different from each other, Fig.4. It is known that the limonene enantiomers cause different odor sensations (R-limonene smells of oranges and S-limonene smells of pine[38]); in essence, the tactoids visualize different scents of the R- and S-limonenes. Aqueous nature of LCLCs solutions adds another useful feature for the amplification and detection of those chiral organic molecules that are soluble in water[39]. Note, however, that the sense of the director twist in the tactoids should not be understood as the direct function of the chiral molecule's conformation; it is known that the same chiral additives might change the value and even the sign of the helical twisting power in the liquid crystalline environment as a function, of, say, temperature or wavelength, see, for example, [40]. Thus the tactoids could distinguish the two enantiometric states only when probed under the same conditions.

The spontaneously twisted tactoids are not the only type of LCLC systems that can be used for detection of chiral additives. Spontaneously twisted structures of LCLC have been reported recently in capillaries with different types of surface anchoring[13, 41-43] and around colloidal spheres dispersed in LCLCs[44]. These structures are also expected to demonstrate a broken parity of left- and right-twists in presence of chiral molecules or particles and thus can be potentially used for the chiral amplification and detection, similarly to the tactoids. This might represent an interesting step in future studies. Other open question worthy of exploration is the effect of partitioning of chiral additives between the nematic and isotropic phases in the biphasic region. Of interest would be also to explore how different enantiomers, say R- and S-forms of limonene, would modify the textures of tactoids is added in sequence, and to what extent the induced changes are reversible. Finally, the very mechanism by which the chiral dopants cause the chromonic aggregates and the director to twist in space, remains to be understood. It is not clear to what extent the director twist depends on the ability of the chiral molecules to intercalate within the aggregates or to adsorb at their surfaces or to change the environment (electric double layers and water structure) around the chromonic aggregates.

To conclude, we demonstrate that minute quantities of chiral molecules such as amino acid L-alanine, limonene and brucine added to the biphasic system with the LCLC tactoids surrounded by isotropic melt, lead to chiral amplification characterized by an increase of optical activity by a factor of $10^3$–$10^4$ as compared to isotropic melts of the same chemical composition. The mechanism is rooted in the long-range elastic nature of orientational order in LCLCs and twist deformations within the tactoids that exists even without the chiral additives. The distinct advantages of direct optical visualization of presence of different enantiomers, high amplification factor, aqueous and non-toxic character of the LCLC systems make the proposed approach of potential practical interest.

The study also poses many questions on the concrete mechanisms of chirality transfer from the molecular to the macroscopic scale in LCLCs; the issue is currently being addressed for a large number of other organic systems in a growing number of studies [45-50]. Note in this regard that the aggregation of DNA base pairs is, to some extent, similar to the aggregation of LCLC molecules[14,51]. Clark and co-workers found that the DNA duplex oligomers can form nematic tactoids in the isotropic surrounding in the mixture of single-stranded and double-stranded oligomers[52, 53]. In this DNA system, one might expect an interplay of chemically and physically (confinement) induced twists similar to the effect described in our paper.

## Acknowledgements

This work was supported by NSF grant DMS-1435372. The authors acknowledge fruitful discussions with P. Collings and R. Selinger. We thank A. Sharma and T. Hegmann for the help with CD measurements.

## Notes and references

[a] Liquid Crystal Institute and Chemical Physics Interdisciplinary Program, Kent State University, Kent, Ohio 44242, USA. *Fax: +1 33 0672 4844; Tel:+1 33 0672 4844; E-mail: olavrent@kent.edu*
† Electronic supplementary information (ESI) available: Circular dichroism measurements of isotropic LCLC solutions with chiral additives.

1. K. Soai, T. Shibata, H. Morioka and K. Choji, *Nature*, 1995, **378**, 767-768.
2. T. Kawasaki, Y. Matsumura, T. Tsutsumi, K. Suzuki, M. Ito and K. Soai, *Science*, 2009, **324**, 492-495.
3. M. M. Green, N. C. Peterson, T. Sato, A. Teramoto, R. Cook and S. Lifson, *Science*, 1995, **268**, 1860-1866.
4. R. Nonokawa and E. Yashima, *Journal of the American Chemical Society*, 2002, **125**, 1278-1283.
5. A. R. A. Palmans and E. W. Meijer, *Angewandte Chemie International Edition*, 2007, **46**, 8948-8968.
6. L. J. Prins, F. De Jong, P. Timmerman and D. N. Reinhoudt, *Nature*, 2000, **408**, 181-184.
7. L. J. Prins, J. Huskens, F. De Jong, P. Timmerman and D. N. Reinhoudt, *Nature*, 1999, **398**, 498-502.
8. R. Fasel, M. Parschau and K.-H. Ernst, *Nature*, 2006, **439**, 449-452.
9. M. Lahav, *Origins of Life and Evolution of Biospheres*, 2007, **37**, 371-377.
10. J. Lydon, *Liq Cryst*, 2011, **38**, 1663-1681.
11. S. W. Tam-Chang and L. M. Huang, *Chem Commun*, 2008, DOI: Doi 10.1039/B714319b, 1957-1967.
12. P. J. Collings, A. J. Dickinson and E. C. Smith, *Liq Cryst*, 2010, **37**, 701-710.
13. J. Jeong, Z. S. Davidson, P. J. Collings, T. C. Lubensky and A. G. Yodh, *Proceedings of the National Academy of Sciences*, 2014, **111**, 1742-1747.
14. H.-S. Park and O. Lavrentovich, in *Liquid Crystals Beyond Displays: Chemistry, Physics, and Applications*, John Wiley & Sons, Inc. , 2012, ch. 14, pp. 449-484.
15. Z. Dogic, *Physical Review Letters*, 2003, **91**, 165701.
16. N. Puech, E. Grelet, P. Poulin, C. Blanc and P. van der Schoot, *Physical Review E*, 2010, **82**, 020702.

17. L. Tortora and O. D. Lavrentovich, *Proceedings of the National Academy of Sciences*, 2011, **108**, 5163-5168.
18. Y.-K. Kim, S. V. Shiyanovskii and O. D. Lavrentovich, *Journal of Physics: Condensed Matter*, 2013, **25**, 404202.
19. S. Zhou, Y. A. Nastishin, M. M. Omelchenko, L. Tortora, V. G. Nazarenko, O. P. Boiko, T. Ostapenko, T. Hu, C. C. Almasan, S. N. Sprunt, J. T. Gleeson and O. D. Lavrentovich, *Phys Rev Lett*, 2012, **109**, 037801.
20. S. Zhou, K. Neupane, Y. A. Nastishin, A. R. Baldwin, S. V. Shiyanovskii, O. D. Lavrentovich and S. Sprunt, *Soft Matter*, 2014, **10**, 6571-6581.
21. W. Thiemann and H. Teutsch, *Origins of Life and Evolution of the Biosphere*, 1990, **20**, 121-126.
22. Y. A. Nastishin, H. Liu, T. Schneider, V. Nazarenko, R. Vasyuta, S. V. Shiyanovskii and O. D. Lavrentovich, *Physical Review E*, 2005, **72**, 041711.
23. P. Yeh and C. Gu, in *Optics of liquid crystal displays*, Wiley-Interscience, New York, 1999, pp. 119-130.
24. L. D. Barron, *Molecular light scattering and optical activity*, Cambridge University Press, 2004.
25. I. Blei and G. Odian, *General, Organic, and Biochemistry: Connecting Chemistry to Your Life*, W. H. Freeman and Company, 2000.
26. G. Volovik and O. Lavrentovich, *Zhurn. Eksp. Teor. Fiz*, 1983, **85**, 1997-2010.
27. O. Lavrentovich and V. Sergan, *IL Nuovo Cimento D*, 1990, **12**, 1219-1222.
28. C. K. McGinn, L. I. Laderman, N. Zimmermann, H.-S. Kitzerow and P. J. Collings, *Physical Review E*, 2013, **88**, 062513.
29. Y. Wang, J. Xu, Y. Wang and H. Chen, *Chemical Society Reviews*, 2013, **42**, 2930-2962.
30. L. J. Prins, P. Timmerman and D. N. Reinhoudt, *Journal of the American Chemical Society*, 2001, **123**, 10153-10163.
31. M. M. Green, M. P. Reidy, R. D. Johnson, G. Darling, D. J. O'Leary and G. Willson, *Journal of the American Chemical Society*, 1989, **111**, 6452-6454.
32. A. R. A. Palmans, J. A. J. M. Vekemans, E. E. Havinga and E. W. Meijer, *Angewandte Chemie International Edition in English*, 1997, **36**, 2648-2651.
33. L. Brunsfeld, B. G. G. Lohmeijer, J. A. J. M. Vekemans and E. W. Meijer, *Journal of Inclusion Phenomena and Macrocyclic Chemistry*, 2001, **41**, 61-64.
34. M. M. Green, B. A. Garetz, B. Munoz, H. Chang, S. Hoke and R. G. Cooks, *Journal of the American Chemical Society*, 1995, **117**, 4181-4182.
35. J. van Gestel, A. R. A. Palmans, B. Titulaer, J. A. J. M. Vekemans and E. W. Meijer, *Journal of the American Chemical Society*, 2005, **127**, 5490-5494.
36. H. Fenniri, B.-L. Deng and A. E. Ribbe, *Journal of the American Chemical Society*, 2002, **124**, 11064-11072.
37. J. van Gestel, *Macromolecules*, 2004, **37**, 3894-3898.
38. C. McManus, *Right hand, left hand: The origins of asymmetry in brains, bodies, atoms and cultures*, Harvard University Press, Cambridge, MA, 2002.
39. E. Yashima, K. Maeda and T. Nishimura, *Chemistry – A European Journal*, 2004, **10**, 42-51.
40. H.-G. Kuball and H. Brüning, *Chirality*, 1997, **9**, 407-423.
41. J. Jeong, L. Kang, Z. S. Davidson, P. J. Collings, T. C. Lubensky and A. G. Yodh, *Proceedings of the National Academy of Sciences*, 2015, **112**, E1837-E1844.
42. Z. S. Davidson, L. Kang, J. Jeong, T. Still, P. J. Collings, T. C. Lubensky and A. G. Yodh, *Physical Review E*, 2015, **91**, 050501.
43. R. Chang, K. Nayani, J. Fu, E. Reichmanis, J. O. Park and M. Srinivasarao, *Bulletin of the American Physical Society*, 2015, **60**, P1.00033.
44. A. Nych, U. Ognysta, I. Muševič, D. Seč, M. Ravnik and S. Žumer, *Physical Review E*, 2014, **89**, 062502.
45. A. Harris, R. D. Kamien and T. Lubensky, *Reviews of Modern Physics*, 1999, **71**, 1745.
46. N. Petit-Garrido, J. Ignés-Mullol, J. Claret and F. Sagués, *Physical Review Letters*, 2009, **103**, 237802.
47. N. Petit-Garrido, J. Claret, J. Ignés-Mullol and F. Sagués, *Nature Communications*, 2012, **3**, 1001.

48. N. Micali, H. Engelkamp, P. G. van Rhee, P. C. M. Christianen, L. M. Scolaro and J. C. Maan, *Nature Chemistry*, 2012, **4**, 201-207.
49. C. Wattanakit, Y. B. Saint Côme, V. Lapeyre, P. A. Bopp, M. Heim, S. Yadnum, S. Nokbin, C. Warakulwit, J. Limtrakul and A. Kuhn, *Nature Communications*, 2014, **5**, 3325.
50. T. Ohzono, T. Yamamoto and J.-i. Fukuda, *Nature Communications*, 2014, **5**, 3735.
51. K. Mundy, J. Sleep and J. Lydon, *Liquid Crystals*, 1995, **19**, 107-112.
52. M. Nakata, G. Zanchetta, B. D. Chapman, C. D. Jones, J. O. Cross, R. Pindak, T. Bellini and N. A. Clark, *Science*, 2007, **318**, 1276-1279.
53. G. Zanchetta, M. Nakata, M. Buscaglia, T. Bellini and N. A. Clark, *Proceedings of the National Academy of Sciences*, 2008, **105**, 1111-1117.